\documentclass[12pt,twoside,prd,amsmath,amssymb,tightenlines,showkeys,
showpacs,eqsecnum]{revtex4}
\usepackage{epsfig}
\usepackage{mathptmx}
\usepackage{hyperref}
\usepackage{bm}
\usepackage{graphicx}

\renewcommand{\cal}{\mathcal}

\newcommand {\ve}{\varepsilon}
\newcommand {\pr}{\partial}
\newcommand {\cG}{\cal G}
\newcommand {\cD}{\cal D}
\newcommand {\cL}{\cal L}
\newcommand {\G}{\Gamma}
\newcommand {\bg}{\bar \gamma}
\newcommand {\bp}{\bar \psi}
\newcommand {\vf}{\varphi}
\newcommand {\bv}{\bar v}
\newcommand {\p}{\psi}

\def \myfigures #1#2#3#4#5#6#7#8
{\begin{figure}[ht]
    \begin{center}
        \includegraphics[width=#2 \textwidth]{#1.eps}
        \hfill
        \includegraphics[width=#6 \textwidth]{#5.eps}
        \parbox[t]{#4\textwidth}{\caption {#3}\label{#1}}
        \hfill
        \parbox[t]{#8\textwidth}{\caption {#7}\label{#5}}
    \end{center}
\end{figure} }

\def \myfigs #1#2#3#4#5#6#7#8
{\begin{figure}[ht]
    \begin{center}
        \includegraphics[width=#2 \textwidth, angle=-90]{#1.eps}
        \hfill
        \includegraphics[width=#6 \textwidth, angle=-90]{#5.eps}
        \vskip 5 mm
        \parbox[t]{#4\textwidth}{\caption {#3}\label{#1}}
        \hfill
        \parbox[t]{#8\textwidth}{\caption {#7}\label{#5}}
    \end{center}
\end{figure}}

\def \myfigss #1#2#3#4#5#6#7#8
{\begin{figure}[ht]
    \begin{center}
        \includegraphics[width=#2 \textwidth, angle=-90]{#1.eps}
        \hfill
        \includegraphics[width=#6 \textwidth]{#5.eps}
        \vskip 5 mm
        \parbox[t]{#4\textwidth}{\caption {#3}\label{#1}}
        \hfill
        \parbox[t]{#8\textwidth}{\caption {#7}\label{#5}}
    \end{center}
\end{figure}}

\def\myfigure #1#2#3#4
{\begin{figure}[ht]\begin{center}
\includegraphics[width=#2 \textwidth]{#1.eps}
\parbox[t]{#4\textwidth}{\caption{#3}\label{#1}}
\end{center}\end{figure}}

\begin{document}
\title{Interacting spinor and scalar fields in Bianchi
cosmology}
\author{Bijan Saha}
\affiliation{Laboratory of Information Technologies\\
Joint Institute for Nuclear Research, Dubna\\
141980 Dubna, Moscow region, Russia} \email{bijan@jinr.ru}
\homepage{http://www.jinr.ru/~bijan/}
\date{\today}
\begin{abstract}
A self-consistent system of interacting spinor and scalar fields
is considered within the scope of Bianchi type VI cosmological
model filled with a perfect fluid. The contribution of the
cosmological constant ($\Lambda$-term) is taken into account as
well. Exact self-consistent solutions to the field equations are
obtained for a special choice of spatial inhomogeneity and the
interaction terms of the spinor and scalar fields. It has been found
that some special choice of metric functions can give rise to a
singularity-free solutions independent of the value and sign of
the $\Lambda$ term. It is also shown that the introduction of a
positive $\Lambda$, the most widespread kind of dark energy, leads
to the rapid growth of the universe, while the negative one,
corresponding to an additional gravitational energy gives rise to
an oscillatory or non-periodic mode of expansion. The role of the
spatial inhomogeneity in the evolution of the universe is
clarified within the scope of the considered models.
\end{abstract}



\maketitle
\bigskip


\section{Introduction}
\setcounter{equation}{0}

The role of the spinor field in the evolution of the Universe has
been studied by a number of authors
\cite{sactp1,sagrg,sajmp,sal,saprd,bited,green}. A principal goal
of those studies was to find out singularity-free solutions to the
corresponding field equations. The gravitational field in these
cases was given by an anisotropic Bianchi type-I (BI) cosmological
model. It has been found that the introduction of the nonlinear spinor
field in some special cases results in a rapid growth of the
Universe. In light of that study many authors believe it
is possible to consider the spinor field as one of the candidates
to explain the late time accelerated mode of expansion
\cite{kremer1,sstan,soprd}.

Though as an anisotropic cosmological model cosmologists basically consider
Bianchi type-I space-time, there are still a few other models that
describe an anisotropic space-time and generate particular interest among
physicists~ \cite{singh,weaver,coley,medina,berger,aposto,papado,chris}. In
Ref. \cite{weaver} methods of dynamical systems analysis were used to show that
the presence of a magnetic field orthogonal to the two commuting Killing vector
fields in any spatially homogeneous Bianchi type $VI_0$ vacuum solution to
Einstein's equation changes the evolution towards the singularity from collapse
to bounce. The authors in Ref. \cite{coley} studied the problem of
isotropization of scalar field Bianchi models with an exponential potential(s).
Other papers mentioned above are devoted to tilted perfect fluid solutions,
chaotic singularities, and conditional symmetries.

In a recent paper \cite{prdbvi} we studied the self-consistent
system of the nonlinear spinor field and an anisotropic
inhomogeneous gravitational field in order to clarify the role of
the spinor field nonlinearity and the space-time inhomogeneity in
the formation of a singularity-free universe. As an anisotropic
space-time we chose a Bianchi type-VI (BVI) model, since a
suitable choice of its parameters yields a few other Bianchi
models including BI and FRW universes. It can be noted that unlike
the BI universe, the BVI space-time is inhomogeneous. Inclusion of
inhomogeneity in the gravitational field significantly complicates
the search for an exact solution to the system. The purpose of
this paper is to study a self consistent system of spinor,
scalar and Bianchi type-VI gravitational fields in presence of a
perfect fluid and a cosmological constant and study the role of
corresponding material fields in the evolution of the universe. \noindent

\vskip 5mm
\section{Basic equations and their general solutions}

We shall investigate a self-consistent system of nonlinear spinor
and Einstein gravitational fields. These two fields are to be
determined by the following action:
\begin{equation}
{\cal S}({\rm g};\p,\bp) = \int \cL \sqrt{-{\rm g}}{\rm d \Omega}
\label{action}
\end{equation}
with
\begin{equation}
{\cL} = {\cL}_{\rm g} + {\cL}_{\rm sp} + {\cL}_{\rm sc} +
{\cL}_{\rm int} + {\cL}_{\rm pf}. \label{lag}
\end{equation}
The gravitational part of the Lagrangian \eqref{lag} ${\cL}_{\rm
g}$ is given by a Bianchi type-VI (BVI hereafter) space-time,
whereas the terms ${\cL}_{\rm sp},\,{\cL}_{\rm sc}$, and
${\cL}_{\rm int}$ describe the spinor and scalar field Lagrangian
and an interaction between them, respectively. The term $
{\cL}_{\rm pf}$ describes the Lagrangian density of the perfect
fluid which minimally couples to the spinor and scalar fields
through gravitational one.

             \subsection{Matter field Lagrangian}

For a spinor field $\p$, the symmetry between $\p$ and $\bp$
appears to demand that one should choose the symmetrized
Lagrangian~ \cite{kibble}. Keeping this in mind we choose the
spinor field Lagrangian as
\begin{equation}
{\cal L}_{sp}=\frac{i}{2} \biggl[ \bp \gamma^{\mu} \nabla_{\mu}
\p- \nabla_{\mu} \bp \gamma^{\mu} \p \biggr] - M\bp \p.
\label{lspin}
\end{equation}
Here $M$ is the spinor mass, $\nabla_\mu$ is the covariant
derivatives acting on a spinor field as ~\cite{Zelnor,brill}
\begin{equation}
\label{cvd} \nabla_\mu \psi = \frac{\partial \psi}{\partial x^\mu}
-\G_\mu \psi, \quad \nabla_\mu \bp = \frac{\partial \bp}{\partial
x^\mu} + \bp \G_\mu,
\end{equation}
where $\G_\mu$ are the Fock-Ivanenko spinor connection
coefficients defined by
\begin{equation}
\G_\mu = \frac{1}{4} \gamma^\sigma \Bigl(\G_{\mu \sigma}^{\nu}
\gamma_{\nu} - \partial_{\mu} \gamma_{\sigma}\Bigr). \label{fock}
\end{equation}

The massless scalar field Lagrangian is chosen to be
\begin{equation}
{\cal L}_{\rm sc} = \frac{1}{2} \vf_{,\alpha}\vf^{,\alpha}.
\label{lsc}
\end{equation}

The interaction between the spinor and scalar fields is given by
the Lagrangian~\cite{sagrg}
\begin{equation}
{\cal L}_{\rm int}= \frac{\lambda}{2}\,\vf_{,\alpha}\vf^{,\alpha}
F. \label{lint}
\end{equation}
Here $\lambda$ is the self-coupling
constant and $F = F(I,J)$ is some arbitrary functions of
invariants $I = S^2$ and $J = P^2$ generated from the real
bilinear forms $S= \bar \psi \psi$ and $P = i \bar \psi \gamma^5
\psi$ of the spinor field.

The contribution of the perfect fluid to the system is performed
by means of its energy-momentum tensor, which acts as one of the
sources of the corresponding gravitational field equations. So
here we do not need to write the Lagrangian density ${\cL}_{\rm
pf}$ explicitly. The reason for writing ${\cL}_{\rm pf}$ in Eqs.
\eqref{action} and \eqref{lag} is to underline that we are dealing
with a self-consistent system. An interesting discussion on the
action and Lagrangian for a perfect fluid can be found in Refs.
\cite{dirac,milekhin1,milekhin2}.

\subsection{The gravitational field}

The gravitational part of the Lagrangian in \eqref{lag} has the
form:
\begin{equation}
\cL_{\rm grav} = \frac{R}{2 \kappa}, \label{gr}
\end{equation}
Here $R$ is the scalar curvature and $\kappa$ is Einstein's gravitational
constant. The gravitational field in our case is given by a BVI metric:
\begin{equation}
ds^2 = dt^2 - a^{2} e^{-2mz}\,dx^{2} - b^{2} e^{2nz}\,dy^{2} -
c^{2}\,dz^2,
\label{bvi}
\end{equation}
with $a,\,b,\,c$ being functions of time only. Here $m,\,n$ are
some arbitrary constants and the velocity of light is taken to be
unity. It should be emphasized that the BVI metric models a
universe that is anisotropic and inhomogeneous. A suitable choice
of $m,\,n$ as well as the metric functions $a,\,b,\,c$ in the BVI
metric given by \eqref{bvi} generates the following Bianchi-type
universes: (i) for $m=n$ the BVI metric transforms into a
Bianchi-type V (BV) universe; (ii) for $n=0$ the BVI metric
transforms into a Bianchi-type III (BIII) universe; (iii) for $m=n
=0$ the BVI metric transforms into a Bianchi-type I (BI) universe
and finally, (iv) for $m=n=0$ and an equal scale factor in all
three directions the BVI metric transforms into a FRW universe.

The metric \eqref{bvi} has the following nontrivial components of
Riemann and Ricci tensors:
\begin{eqnarray}
R_{\,\,\,01}^{01} &=& -\frac{\ddot a}{a}, \quad
R_{\,\,\,02}^{02} = -\frac{\ddot b}{b}, \quad
R_{\,\,\,03}^{03} = -\frac{\ddot c}{c},
\nonumber \\
R_{\,\,\,12}^{12} &=& -\frac{mn}{c^2} -\frac{\dot a}{a}\frac{\dot
b}{b}, \quad R_{\,\,\,13}^{13} = \frac{m^2}{c^2} - \frac{\dot
c}{c}\frac{\dot a}{a}, \quad R_{\,\,\,23}^{23} = \frac{n^2}{c^2}
-\frac{\dot b}{b}\frac{\dot c}{c}, \nonumber
\end{eqnarray}
\begin{eqnarray}
R_{3}^{0} &=& \Bigl(m\frac{\dot a}{a} - n\frac{\dot b}{b} - (m -n ) \frac{\dot
c}{c}\Bigr), \nonumber\\
R_{0}^{0} &=& -\Bigl(\frac{\ddot a}{a^2} + \frac{\ddot b}{b^2} + \frac{\ddot
c}{c^2}\Bigr), \nonumber\\
R_{1}^{1} &=& - \Bigl(\frac{\ddot a}{a}+ \frac{\dot a}{a}\frac{\dot b}{b} +
\frac{\dot a}{a}\frac{\dot c}{c} -\frac{m^2 - mn}{c^2}\Bigr), \nonumber\\
R_{2}^{2} &=& - \Bigl(\frac{\ddot b}{b}+ \frac{\dot a}{a}\frac{\dot b}{b} +
\frac{\dot b}{b}\frac{\dot c}{c} -\frac{n^2 - mn}{c^2}\Bigr), \nonumber\\
R_{3}^{3} &=& - \Bigl(\frac{\ddot c}{c}+ \frac{\dot a}{a}\frac{\dot c}{c} +
\frac{\dot b}{b}\frac{\dot c}{c} -\frac{m^2 + n^2}{c^2}\Bigr). \nonumber
\end{eqnarray}
We write the components of the Riemann and Ricci tensors as 
invariant characteristics of space-time which one needs to know
in order to investigate the existence of a singularity (singular
point) are composed of these tensors together with the metric
tensor. Although in 4D Riemann space there are 14 independent
invariants \cite{mits,saprd}, it is sufficient to study only three
of them, namely the scalar curvature $I_1 = R$, $I_2 = R_{\mu
\nu}R^{\mu\nu}$ and the Kretschmann scalar $I_3 = R_{\alpha \beta
\mu \nu} R^{\alpha \beta \mu \nu}$ \cite{bronshik,fay}. From the
Riemann and Ricci tensors written above one finds
\begin{subequations}
\label{gravinv}
\begin{eqnarray}
I_1 &=& R = -\frac{2}{\tau}\biggl[{\ddot \tau}-{\dot a}{\dot b}c -a {\dot b}
{\dot c}-{\dot a}b{\dot c}- \frac{a b}{c}(m^2 - mn + n^2)
\biggr],\\
I_2 &=& \bigl(R_0^0\bigr)^2 + \bigl(R_1^1\bigr)^2 + \bigl(R_2^2\bigr)^2
+ \bigl(R_3^3\bigr)^2+ R_3^0 R_0^3,\\
I_3 &=& 4 [ \bigl(R_{\,\,\,01}^{01}\bigr)^2 + \bigl(R_{\,\,\,02}^{02}\bigr)^2 +
\bigl(R_{\,\,\,03}^{03}\bigr)^2 + \bigl(R_{\,\,\,12}^{12}\bigr)^2 +
\bigl(R_{\,\,\,31}^{31}\bigr)^2 + \bigl(R_{\,\,\,23}^{23}\bigr)^2 ],
\end{eqnarray}
\end{subequations}
where we define
\begin{equation}
\tau = a b c.
\label{taudef}
\end{equation}
From \eqref{gravinv} it follows that $I_1 \propto 1/\tau$, $I_2
\propto 1/\tau^2$, and $I_3 \propto 1/\tau^2$. Note that the
remaining 11 invariants are composed of two or more Ricci and/or
Riemann tensors and hence are inversely proportional to
$(\tau)^r$, where $r$ is the number of tensors in the
corresponding invariant. Thus we see that at any space-time point
where $\tau = 0$, the invariants $I_1, I_2, I_3$ become infinity;
hence the space-time becomes singular at this point.

\subsection{Field equations}

Let us now write the field equations corresponding to the action
\eqref{action}.

Variation of Eq. \eqref{action} with respect to the spinor field
$\psi\,(\bp)$ gives the following spinor field equations:
\begin{subequations}
\label{speq}
\begin{eqnarray}
i\gamma^\mu \nabla_\mu \psi - M \psi + {\cD} \psi +
{\cG} i \gamma^5 \psi &=&0, \label{speq1} \\
i \nabla_\mu \bp \gamma^\mu +  M \bp - {\cD} \bp - {\cG} i \bp
\gamma^5 &=& 0, \label{speq2}
\end{eqnarray}
\end{subequations}
where we use the notation
$$ {\cD} =  \lambda S \vf_{,\alpha}\vf^{,\alpha} \frac{\pr F}{\pr I} =
\frac{\lambda}{2} \vf_{,\alpha}\vf^{,\alpha} \frac{\pr F}{\pr S} ,
\quad {\cG} =  \lambda P \vf_{,\alpha}\vf^{,\alpha} \frac{\pr
F}{\pr J} = \frac{\lambda}{2} \vf_{,\alpha}\vf^{,\alpha} \frac{\pr
F}{\pr P}.$$ Since the nonlinearity in the foregoing equations is
generated by the interacting scalar field, Eqs. \eqref{speq} can
be viewed as spinor field equations with induced nonlinearity.

Variation of Eq. \eqref{action} with respect to the scalar field
yields the following scalar field equation:
\begin{equation}
\frac{1}{\sqrt{-g}} \frac{\pr}{\pr x^\nu} \Bigl(\sqrt{-g}
g^{\nu\mu} (1 + \lambda F) \vf_{,\mu}\Bigr) = 0. \label{scfe}
\end{equation}

Finally, varying Eq. \eqref{action} with respect to metric tensor
$g_{\mu\nu}$ one finds the Einstein's field equations. On account
of the $\Lambda$ term they have the form
\begin{equation}
R_{\mu}^{\nu} - \frac{1}{2} \delta_{\mu}^{\nu} R = \kappa
T_{\mu}^{\nu} + \delta_{\mu}^{\nu} \Lambda. \label{eing}
\end{equation}
where $R_{\nu}^{\mu}$ is the Ricci tensor, $R$ is the Ricci
scalar, and $T_{\nu}^{\mu}$ is the energy-momentum tensor of the
matter fields. In our case, where space-time is given by a BVI
metric \eqref{bvi}, the equations for the metric functions
$a,\,b,\,c$ read
\begin{subequations}
\label{ein}
\begin{eqnarray}
\frac{\ddot b}{b} +\frac{\ddot c}{c} +\frac{\dot b}{b}\frac{\dot
c}{c} - \frac{n^2}{c^2} &=& \kappa T_{1}^{1} + \Lambda, \label{11}\\
\frac{\ddot c}{c} +\frac{\ddot a}{a} +\frac{\dot c}{c}\frac{\dot
a}{a} - \frac{m^2}{c^2} &=& \kappa T_{2}^{2} + \Lambda, \label{22} \\
\frac{\ddot a}{a} +\frac{\ddot b}{b} +\frac{\dot a}{a}\frac{\dot
b}{b} + \frac{m n}{c^2} &=& \kappa T_{3}^{3} + \Lambda, \label{33}\\
\frac{\dot a}{a}\frac{\dot b}{b} +\frac{\dot b}{b}
\frac{\dot c}{c} +
\frac{\dot c}{c}\frac{\dot a}{a} - \frac{m^2 - m n + n^2}{c^2} &=&
\kappa T_{0}^{0} + \Lambda, \label{00}\\
m \frac{\dot a}{a} - n \frac{\dot b}{b}
- (m - n) \frac{\dot c}{c} &=& \kappa T_{3}^{0}. \label{03}
\end{eqnarray}
\end{subequations}
Here over dots denote differentiation with respect to time ($t$).
The energy-momentum tensor of the material field $T_{\mu}^{\nu}$
is given by
\begin{equation}
T_{\mu}^{\nu} = T_{\mu {\rm (sp)}}^{\,\,\,\nu} + T_{\mu {\rm
(sc)}}^{\,\,\,\nu} + T_{\mu {\rm (int)}}^{\,\,\,\nu} + T_{\mu {\rm
(pf)}}^{\,\,\,\nu}. \label{tem}
\end{equation}
Here $T_{\mu {\rm (sp)}}^{\,\,\,\nu}$ is the energy momentum
tensor of the spinor field defined by
\begin{equation}
T_{\mu {\rm (sp)}}^{\,\,\,\rho}=\frac{i}{4} g^{\rho\nu} \biggl(\bp
\gamma_\mu \nabla_\nu \psi + \bp \gamma_\nu \nabla_\mu \psi -
\nabla_\mu \bar \psi \gamma_\nu \psi - \nabla_\nu \bp \gamma_\mu
\psi \biggr) \,- \delta_{\mu}^{\rho}{\cal L}_{\rm sp}.
\label{temsp}
\end{equation}
The term  ${\cal L}_{\rm sp}$ in view of Eq. \eqref{speq} takes
the form
\begin{equation}
{\cal L}_{\rm sp} = -\bigl({\cD} S + {\cG} P\bigr). \label{lsp}
\end{equation}
The energy momentum tensor of the scalar field is given by
\begin{equation}
T_{\mu {\rm (sc)}}^{\,\,\,\nu}= \vf_{,\mu}\vf^{,\nu} -
\delta_{\mu}^{\nu}{\cal L}_{\rm sc}. \label{temsc}
\end{equation}
For the interaction field we find
\begin{equation}
T_{\mu {\rm (int)}}^{\,\,\,\nu}= \lambda F \vf_{,\mu}\vf^{,\nu}
 -  \delta_{\mu}^{\nu} {\cal L}_{\rm int}.
\label{temint}
\end{equation}
$T_{\mu\,{\rm (pf)}}^{\nu}$ is the energy momentum tensor of a
perfect fluid. For a Universe filled with a perfect fluid, in a
comoving system of reference such that $u^\mu = (1,\,0,\,0,\,0)$
we have
\begin{equation}
T_{\mu\,{\rm (pf)}}^{\nu} = (p + \ve) u_\mu u^\nu - \delta_\mu^\nu
p = (\ve,\,-p,\,-p,\,-p).
\end{equation}
The energy $\ve$ and the pressure $p$ of the perfect fluid obey
the following equation of state:
\begin{equation}
p\,=\,\zeta\,\ve, \label{eqst}
\end{equation}
where $\zeta$ is a constant and lies in the interval $\zeta\, \in
[0,\,1]$. Depending on its numerical value, $\zeta$ describes the
following types of Universes \cite{jacobs}: (i) $\zeta = 0$ (dust Universe);
(ii) $\zeta = 1/3$ (radiation Universe); (iii) $\zeta \in (1/3,\,1)$ (hard Universes) and
(iv) $\zeta = 1$ (Zel'dovich Universe or stiff matter).
Here we note once again that the perfect fluid is minimally
coupled to the system. Being one of its sources, the perfect fluid
leaves its trace on the gravitational field which, in turn,
influences the behavior of the spinor and scalar fields.

From \eqref{fock} we find the following spinor connections for the metric \eqref{bvi}
\begin{eqnarray}
\Gamma_0 = 0, \quad \Gamma_1 = \frac{1}{2}\bg^{1}\bigl[{\dot a}
\bg^0 - m \frac{a}{c} \bg^3 \bigr] e^{-mz}, \quad \Gamma_2 =
\frac{1}{2}\bg^{2} \bigl[ {\dot b} \bg^0 + n \frac{b}{c} \bg^3
\bigr] e^{nz}, \quad \Gamma_3 =&\frac{1}{2} {\dot c} \bg^{3}
\bg^0.  \nonumber
\end{eqnarray}
It is easy to show that
$$\gamma^\mu \Gamma_\mu = -\frac{1}{2}\frac{\dot \tau}{\tau}\bg^0
+ \frac{m - n}{2 c}\bg^3.$$

The Dirac matrices $\gamma^\mu(x)$ of the curved space-time are
connected with those of Minkowski space-time as follows:
$$ \gamma^0=\bg^0,\quad \gamma^1 =\bg^1 e^{mz}/a,
\quad \gamma^2=\bg^2 /b e^{nz},\quad \gamma^3 =\bg^3 /c,$$ with
\begin{eqnarray}
\bg^0\,=\,\left(\begin{array}{cc}I&0\\0&-I\end{array}\right), \quad
\bg^i\,=\,\left(\begin{array}{cc}0&\sigma^i\\
-\sigma^i&0\end{array}\right), \quad
\gamma^5 = \bg^5&=&\left(\begin{array}{cc}0&-I\\
-I&0\end{array}\right),\nonumber
\end{eqnarray}
where $\sigma_i$ are the Pauli matrices:
\begin{eqnarray}
\sigma^1\,=\,\left(\begin{array}{cc}0&1\\1&0\end{array}\right),
\quad
\sigma^2\,=\,\left(\begin{array}{cc}0&-i\\i&0\end{array}\right),
\quad
\sigma^3\,=\,\left(\begin{array}{cc}1&0\\0&-1\end{array}\right).
\nonumber
\end{eqnarray}

Let us consider the spinors and the scalars to be functions of $t$
only, i.e.,
\begin{equation}
\p = \p(t), \quad \vf = \vf (t)
\end{equation}
Under this assumption for the the scalar field from \eqref{scfe}
we find
\begin{equation}
\vf = C_{\rm sc} \int \frac{dt}{\tau(1 + \lambda F)},  \quad
C_{\rm sc} = {\rm const.} \label{scfq}
\end{equation}
For the spinor field from \eqref{speq} we obtain
\begin{subequations}
\label{spinv}
\begin{eqnarray}
\bg^0\Bigl(\dot \p + \frac{\dot \tau}{2 \tau} \p\Bigr)
-\Bigl(\frac{m-n}{2 c} \Bigr)\bg^3 \p + i\Phi \p
+ {\cG} \bg^5 \p &=& 0, \\
\Bigl(\dot{\bp} + \frac{\dot \tau}{2 \tau}\bp\Bigr)\bg^0
-\Bigl(\frac{m-n}{2 c} \Bigr) \bp \bg^3 - i\Phi \bp - {\cG} \bp
\bg^5  &=& 0.
\end{eqnarray}
\end{subequations}
Here we define $\Phi = M - {\cD}$.
Let us introduce a new function
$$u_j (t) = \sqrt{\tau} \p_j(t).$$
Then for the components of the nonlinear spinor field from \eqref{spinv}, one
obtains
\begin{subequations}
\label{u}
\begin{eqnarray}
\dot{u}_{1} + i \Phi u_{1} -
\Bigl[\frac{m-n}{2 c} +{\cG}\Bigr] u_{3} &=& 0, \\
\dot{u}_{2} + i \Phi u_{2} +
\Bigl[\frac{m-n}{2 c}  -{\cG}\Bigr] u_{4} &=& 0, \\
\dot{u}_{3} - i \Phi u_{3} -
\Bigl[\frac{m-n}{2 c} -{\cG}\Bigr] u_{1} &=& 0, \\
\dot{u}_{4} - i \Phi u_{4} + \Bigl[\frac{m-n}{2 c} +{\cG}\Bigr]
u_{2} &=& 0.
\end{eqnarray}
\end{subequations}

Using the spinor field equations \eqref{speq} and \eqref{spinv},
it can be shown that the bilinear spinor forms
\begin{eqnarray}
S &=& \bp \p = \bv v, \quad
P = i \bp \bg^5 \p = i \bv \bg^5 v, \quad
A^{0} = \bp \bg^5 \bg^0 \p = \bv \bg^5 \bg^0 v, \nonumber \\
A^{3} &=& \bp \bg^5 \bg^3 \p = \bv \bg^5 \bg^3 v, \quad
V^{0} = \bp \bg^0 \p = \bv \bg^0 v, \quad
V^{3} = \bp \bg^3 \p = \bv \bg^3 v, \nonumber \\
Q^{30} &=& i \bp \bg^3 \bg^0 \p = i \bv \bg^3 \bg^0 v, \quad
Q^{21} = \bp  \bg^0 \bg^3 \bg^5 \p =
i \bp  \bg^2 \bg^1 \p = i \bv  \bg^2 \bg^1 v, \nonumber
\end{eqnarray}
obey the following system of equations:
\begin{subequations}
\label{inv}
\begin{eqnarray}
\dot S_0 - 2 {\cG} A_{0}^{0} &=& 0, \\
\dot P_0 - 2 \Phi A_{0}^{0} &=& 0, \\
\dot A_{0}^{0} -\frac{m-n}{c} A_{0}^{3} + 2 \Phi P_0 + 2 {\cG}
S_0 &=& 0,\\
\dot A_{0}^{3} -\frac{m-n}{c} A_{0}^{0} &=& 0, \\
\dot V_{0}^{0} - \frac{m-n}{c} V_{0}^{3} &=& 0, \\
\dot V_{0}^{3} - \frac{m-n}{c} V_{0}^{0} +
2 \Phi Q_{0}^{30} - 2 {\cG} Q_{0}^{21} &=& 0,\\
\dot Q_{0}^{30} - 2 \Phi V_{0}^{3} &=& 0, \\
\dot Q_{0}^{21} + 2 {\cG} V_{0}^{3} &=& 0,
\end{eqnarray}
\end{subequations}
where we use the notation $F_0 = \tau F$. Combining these
equations and taking the first integral one gets
\begin{subequations}
\begin{eqnarray}
(S_{0})^{2} + (P_{0})^{2} + (A_{0}^{0})^{2} - (A_{0}^{3})^{2} &=&
C_1 = {\rm const.}, \label{I1}\\
(V_{0}^{3})^{2} + (Q_{0}^{30})^{2} + (Q_{0}^{21})^{2} -
(V_{0}^{0})^{2} &=& C_2 = {\rm const.} \label{I2}
\end{eqnarray}
\end{subequations}

Now let us solve the spinor field equations \eqref{u}. From the
first and the third equations of the system \eqref{u} one finds
\begin{equation}
\dot{u}_{13} = ({\cG} - Q) u_{13}^{2} - 2 i \Phi u_{13}
+ ({\cG} + Q),
\label{rik}
\end{equation}
where, we denote $u_{13} = u_1/u_3$ and $Q=[m - n]/2c$. Equation
\eqref{rik} is of the Riccati type~\cite{Kamke} with variable
coefficients. Transformation~\cite{Zaitsev}
\begin{equation}
v_{13} = \exp\Bigl(- \int ({\cG} - Q) u_{13} dt\Bigr),
\end{equation}
leads from the general Riccati equation \eqref{rik} to a second order linear
one, namely,
\begin{equation}
({\cG} -Q) \ddot{v}_{13} + \bigl[ 2 i \Phi ({\cG} - Q) -
\dot{{\cG}} + \dot{Q}\bigr]\dot{v}_{13} + ({\cG} - Q)^2
({\cG} + Q) v_{13} = 0.
\label{v13}
\end{equation}
Sometimes it is easier to solve a linear second order differential equation
than a first order nonlinear equation. Here we give a general solution to
\eqref{rik}. For this purpose we rewrite \eqref{rik} in the form
\begin{equation}
\dot{w}_{13} = ({\cG} - Q) w_{13}^{2} e^{-2i \int \Phi(t) dt}
+ ({\cG} + Q)e^{2i \int \Phi(t) dt},
\label{w13}
\end{equation}
where we set $u_{13} = w_{13} \exp[-2i \int \Phi(t) dt].$ This is an
inhomogeneous nonlinear differential equation for $w_{13}$. The solution for
the homogeneous part of \eqref{w13}, i.e.,
\begin{equation}
\dot{w}_{13} = ({\cG} - Q) w_{13}^{2} \exp \biggl(-2i \int \Phi(t) dt\biggr)
\label{w13h}
\end{equation}
reads
\begin{equation}
w_{13} = -\Biggl[\int ({\cG} - Q) \exp \biggl(-2i \int \Phi(t) dt\biggr) dt + C
\Biggr]^{-1}, \label{solh}
\end{equation}
where $C$ is an arbitrary constant. Then the general solution to the
inhomogeneous Eqn. \eqref{w13} can be presented as
\begin{equation}
w_{13} = -\Biggl[\int ({\cG} - Q) \exp \biggl(-2i \int \Phi(t) dt\biggr) dt +
C(t) \Biggr]^{-1}, \label{solgen}
\end{equation}
with the time dependent parameter $C(t)$ to be determined from
\begin{equation}
\dot{C} = \Biggl[\int ({\cG} - Q) \exp \biggl(-2i \int \Phi(t) dt\biggr) dt +
C(t) \Biggr]^{2} ({\cG} + Q) e^{2i \int \Phi(t) dt}. \label{Cd}
\end{equation}
Thus given a concrete nonlinear term in the Lagrangian and the
solutions of the Einstein equations, one finds the relation between
$u_1$ and $u_3$ ($u_2$ and $u_4$ as well), hence the components
of the spinor field.

Now we study the Einstein equations \eqref{ein}. In doing so, we write the
components of the energy-momentum tensor, which in our case read
\begin{subequations}
\label{cemt}
\begin{eqnarray}
T_{0}^{0}&=& MS  + \frac{1}{2} {\dot \vf}^2 (1 + \lambda F) + \ve, \\
T_{1}^{1} &=& T_{2}^{2} = T_{3}^{3} = {\cD} S + {\cG} P -
\frac{1}{2} {\dot \vf}^2 (1 + \lambda F) - p.
\end{eqnarray}
\end{subequations}
Let us demand the energy-momentum tensor to be conserved, i.e.,
\begin{equation}
T_{\nu;\mu}^{\mu} = T_{\nu,\mu}^{\mu} + \Gamma_{\beta \mu}^{\mu}\,
T_{\nu}^{\beta} - \Gamma_{\nu \mu}^{\beta}\,T_{\beta}^{\mu} = 0. \label{emc}
\end{equation}
Taking into account that $T_{\mu}^{\nu}$ is a function of $t$ only
and $T_1^1 = T_2^2 = T_3^3$, from \eqref{emc} we find
\begin{equation}
\dot {T_0^0} + \frac{\dot \tau}{\tau} \Bigr(T_0^0 - T_1^1) = 0.
\label{empc}
\end{equation}
In view of the scalar field equation and $\Phi \dot S_{0} - {\cG}
\dot P_{0} =  0$ which follows from \eqref{inv} the Eq.
\eqref{empc} yields
\begin{equation}
\dot \ve = - \frac{\dot \tau}{\tau} \Bigl(\ve + p\Bigr).
\label{vep}
\end{equation}
Further using the equation of state (EOS) \eqref{eqst} for $\ve$
and $p$ one finds
\begin{equation}
\ve = \frac{\ve_0}{\tau^{1+\zeta}}, \quad p = \frac{\zeta
\ve_0}{\tau^{1+\zeta}}. \label{vetau}
\end{equation}

Let us return to Eqs. \eqref{ein}. In view of \eqref{cemt}, from \eqref{03}
one obtains the following relation between the metric functions $a,\,b,\,c$:
\begin{equation}
\Bigl(\frac{a}{c}\Bigr)^m = {\cal N} \Bigl(\frac{b}{c}\Bigr)^n,
\quad {\cal N} = {\rm const}. \label{abcrel}
\end{equation}
Subtracting \eqref{11} from \eqref{22} we find
\begin{equation}
\frac{d}{d t}
\Bigl[\tau \frac{d}{dt}\Bigl\{{\rm ln}\Bigl(\frac{a}{b}\Bigr)
\Bigr\}\Bigr] = \frac{m^2 - n^2}{c^2} \tau.
\label{ab1}
\end{equation}
Analogously, subtraction of \eqref{11} from \eqref{33} and \eqref{22} from
\eqref{33} gives
\begin{equation}
\frac{d}{d t} \Bigl[\tau \frac{d}{dt}\Bigl\{{\rm
ln}\Bigl(\frac{a}{c}\Bigr) \Bigr\}\Bigr] = -\frac{mn + n^2}{c^2}
\tau \label{ac1}
\end{equation}
and
\begin{equation}
\frac{d}{d t} \Bigl[\tau \frac{d}{dt}\Bigl\{{\rm
ln}\Bigl(\frac{b}{c}\Bigr) \Bigr\}\Bigr] = -\frac{mn + m^2}{c^2}
\tau, \label{bc1}
\end{equation}
respectively. It can be shown that, in view of \eqref{inv} and \eqref{abcrel},
the Eqs. \eqref{ab1}, \eqref{ac1}, and \eqref{bc1} are interchangeable.

Taking into account that $\tau = a b c$, from \eqref{abcrel} we can write
$a$ and $b$ in terms of $c$, such that
\begin{equation}
a = \biggl[{\cal N}\tau^n c^{m-2n}\biggr]^{1/(m+n)}, \label{atoc}
\end{equation}
and
\begin{equation}
b = \biggl[\tau^m c^{n-2m}/{\cal N}\biggr]^{1/(m+n)}. \label{btoc}
\end{equation}
Thus we find $a$ and $b$ in terms of $c$ and $\tau$. In doing so
we employed only four out of five Einstein equations, leaving
\eqref{00} unused. Addition of \eqref{11}, \eqref{22}, \eqref{33},
and \eqref{00}, multiplied by 3 gives
\begin{equation}
\frac{\ddot \tau}{\tau} = 2 \frac{m^2 - mn + n^2}{c^2} +
\frac{3\kappa}{2}[T_0^0 + T_1^1] + 3 \Lambda,\label{detertau0}
\end{equation}
which in view of \eqref{cemt} takes the form
\begin{equation}
\frac{\ddot \tau}{\tau} = 2 \frac{m^2 - mn + n^2}{c^2} +
\frac{3\kappa}{2}[MS + {\cD} S + {\cG} P +
(1-\zeta)\ve_0/\tau^{1+\zeta}] + 3 \Lambda. \label{detertau}
\end{equation}
As one sees, we only have one equation with two unknowns. In order
to resolve this problem, we have to assume $c$ as a function of
$\tau$ (or vice versa). Given a concrete form of the spinor field
nonlinearity one finds the solution of \eqref{detertau}. This is
exactly what we do in the next section.

\section{Analysis of the result}

In the preceding section we derived equations for the spinor,
scalar and gravitational fields and their general solutions.
Comparing the equation with those in a BI universe (see e.g., Ref.
\cite{saprd,bited}) we conclude that introduction of inhomogeneity
in gravitational (through $m$ and $n$) imposes additional
restriction on the metric functions. In fact, Eq. \eqref{03} which
connects the metric functions $a,\,b,\,c$ among themselves, does
not figure in the BI universe. In the foregoing sections we
obtained the solutions to the field equations in terms of $\tau$,
whereas, the equations for $\tau$ contains the function $c$
explicitly, i.e., we have just one equation for two unknowns
$\tau$ and $c$. In order to resolve this we have to impose some
additional condition relating $\tau$ and $c$. Though this
assumption imposes some restrictions on the metric functions,
though leaving the space-time anisotropic.

Let us consider the case when $F= F(I) = F(S)$ only. In this case
from \eqref{inv} we obtain
\begin{equation}
S = C_0/\tau, \label{S}
\end{equation}
with $C_0$ being some arbitrary integration constant. The
components of the spinor field can be obtained from \eqref{solh}
setting ${\cG} = 0$ and $Q = (m-n)/2c$. As far as $F = F(J)$ is
concerned, the volume scale $\tau$ can be obtained from the
corresponding equations setting spinor mass $M = 0$, while the for
the components of the spinor field we need to set $Q = (m-n)/2c$
and $\Phi = 0$ in \eqref{solh} [In this case from \eqref{inv} one
finds $P = D_0/\tau$, with $D_0$ being some arbitrary integration
constant]. Beside this, we assume $c = c(\tau)$. The Eq.
\eqref{detertau} then can be written as
\begin{equation}
\ddot \tau = {\cal F}(\tau,q), \label{tau2}
\end{equation}
where we define
\begin{equation}
{\cal F}(\tau,q) =  2 (m^2 - mn + n^2)\frac{\tau}{c^2} +
\frac{3\kappa}{2} \bigl[ M + {\cal D} + \frac{1 -
\zeta}{\tau^\zeta}\bigr] + 3 \Lambda \tau. \label{rhs}
\end{equation}
Here $q$ is a set of problem parameters, namely, $q = \{m,\,
n,\,M,\,\lambda,\,\eta,\zeta,\Lambda\}.$ Equation~\eqref{tau2}
admits the following first integral,
\begin{equation}
\dot \tau = \sqrt{2[E - U(\tau,q)]}, \label{velocity}
\end{equation}
with the potential
\begin{equation}
U(\tau,q) = -\{4(m^2 -mn + n^2) \int\frac{\tau d\tau}{c^2} +
3\kappa[M \tau + \int {\cal D}d\tau + \tau^{1 - \zeta}]+3\Lambda
\tau^2 \}. \label{poten}
\end{equation}
From a mechanical point of view, Eq. \eqref{tau2} can be
interpreted as an equation of motion of a single particle with
unit mass under the force $\mathcal F(\tau,q)$. In
\eqref{velocity} $E$ is the integration constant which can be
treated as an energy level, and ${\cal U}(\tau,q)$ is the potential
of the force $\mathcal F(q_1, \tau)$. We solve Eq.
\eqref{tau2} numerically using Runge-Kutta method. The initial
value of $\tau$ is taken to be a reasonably small one, while the
corresponding first derivative $\dot \tau$ is evaluated from
\eqref{velocity} for a given $E$.

In what follows we solve Eq. \eqref{tau2} for (i) $c = \tau$
and (ii) $c = \sqrt{\tau}$.

\subsection{Case with $c = \tau$}

Let us consider this case in details for different types of spinor
analyze field nonlinearity.

\subsubsection{Spinor field with power law induced nonlinearity}

Here we consider the case when the spinor field nonlinearity is
given by a power law of $S$. In doing so we set $F = S^\eta$, with
$\eta$ being the power of nonlinearity. The right hand side of 
Eq. \eqref{tau2} in this case has the form
\begin{eqnarray}
{\cal F}(\tau,q) = 2\frac{m^2 -mn + n^2}{\tau} +
\frac{3\kappa}{2}\biggl[M C_0 + \frac{\lambda C_{\rm s} \eta
\tau^{\eta -1}}{(\tau^\eta + \lambda C_0^\eta)^2}+
\frac{(1-\zeta)\ve_0}{\tau^\zeta}\biggr] + 3 \Lambda \tau,
\label{force1}
\end{eqnarray}
with $C_{\rm s} = C_{\rm sc}^2 C_0^\eta/2$. For the potential in
this case we have
\begin{eqnarray}
{\cal U}(\tau,q) = -\{4(m^2 -mn +n^2) {\rm ln}\,\tau + 3\kappa [M
C_0\,\tau -  \lambda C_{\rm s}/(\tau^\eta + \lambda C_0^\eta) +
\ve_0 \tau^{1-\zeta}] + 3 \Lambda \tau^2\}. \label{poten1}
\end{eqnarray}
We solve Eq. \eqref{tau2} with the right hand side given by
\eqref{force1}. For simplicity further we set $\kappa = 1$, $C_0 =
1$, $C_s = 1$ and $\ve_0 = 1$. For numerical calculations we used
the following values for problem parameters: $m=2$, $n=1$
,$M=1$,$\lambda=0.1$,$\zeta=1/3$,$\eta=3$. Note that $\eta = -3$
gives almost the same result as in case of $\eta = 3$. For
cosmological constants we used $\Lambda = 0, +1, -1$,
respectively. The energy level $E$ is taken to be zero ($E = 0$).
It should be noted that in the present case the potential
possesses an infinitely high barrier at $\tau = 0$, it means in
the case at hand $\tau$ is always positive, that is we have
singularity-free solution. Note that for a given value of $E$ the
minimum value of $\tau$ ($\tau_{\rm min}$) should be greater or
equal to the value of $\tau$ at the point of intersection of $E$
and ${\cal U}$. Here we set $E = 0$, so $\tau_{\rm min} \ge
\tau_{\rm int}$ : ${\cal U} (\tau_{\rm int}) = 0$. For simplicity
here the initial value of $\tau$ is taken to be unity. Moreover,
in case of $\Lambda < 0$ which is responsible for additional
gravitational energy the value of $\tau$ is bound from above as
well, i.e., in this case the value of $\tau$ lies between the two
points of intersection of $E$ and ${\cal U}$.

\myfigures{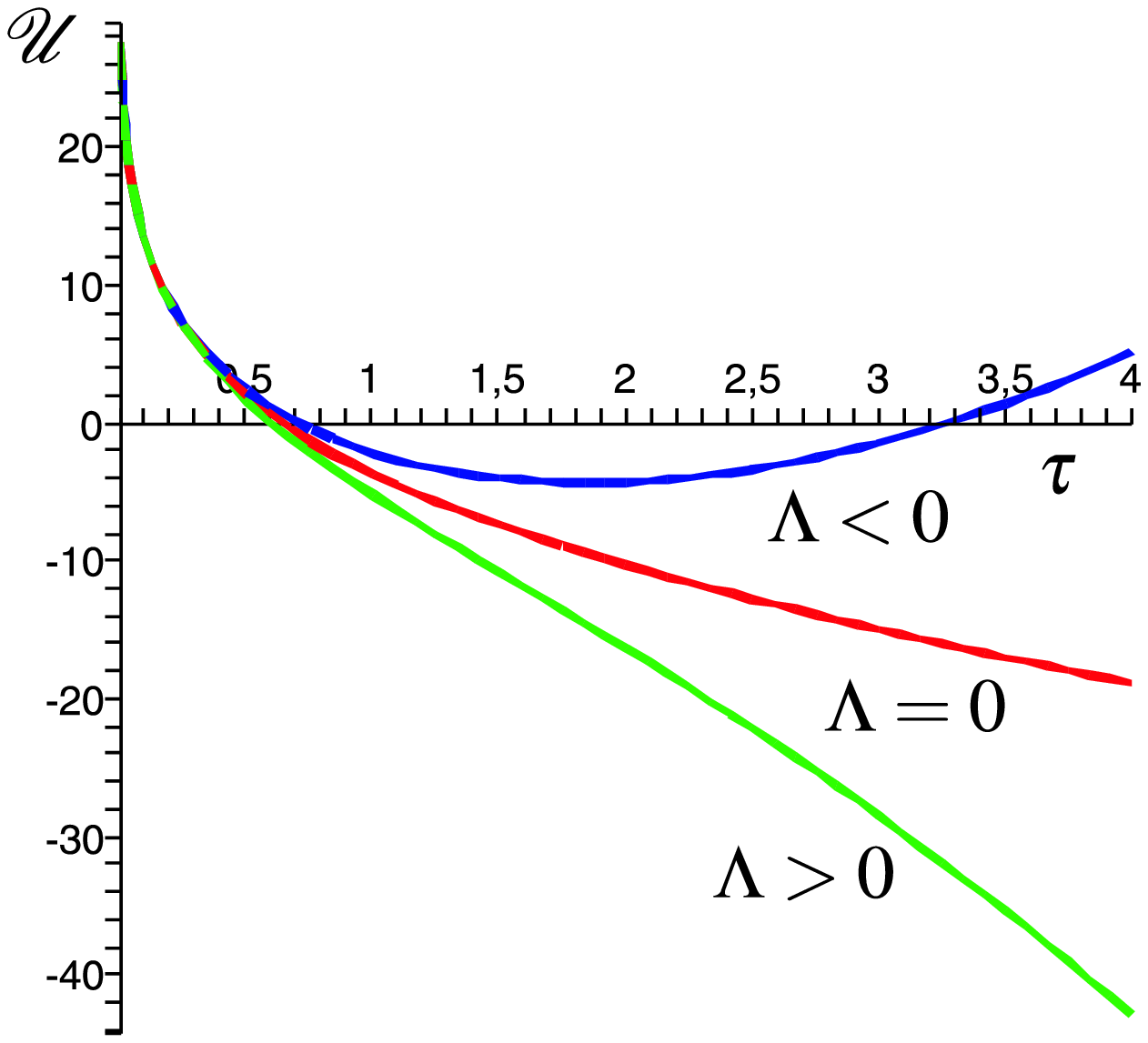}{0.45}{View of the potential ${\cal U}$ given by
\eqref{poten1} for different values of $\Lambda$.}{0.45}
{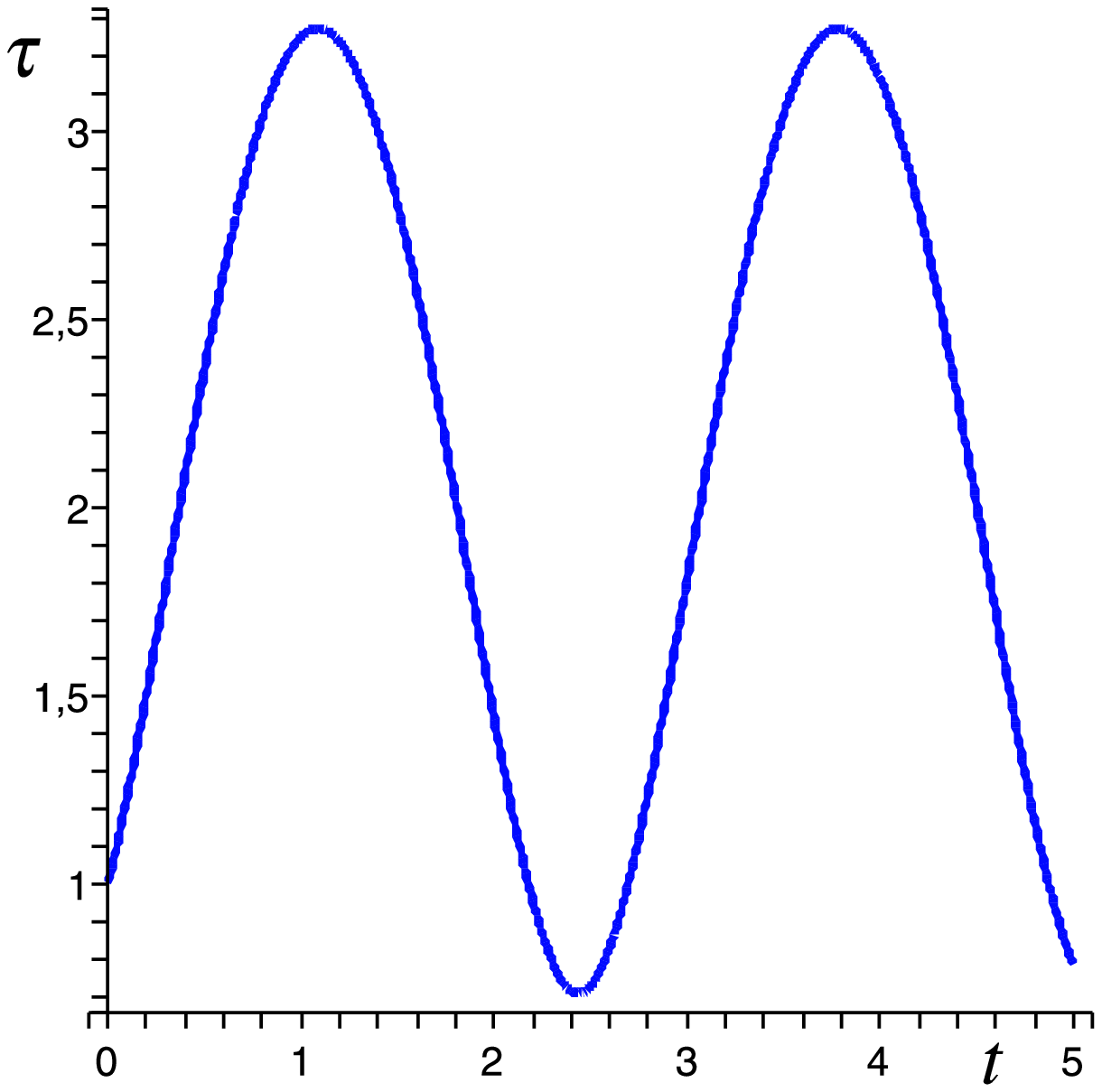}{0.45}{Evolution of $\tau$ for a negative $\Lambda$.}{0.45}

\myfigures{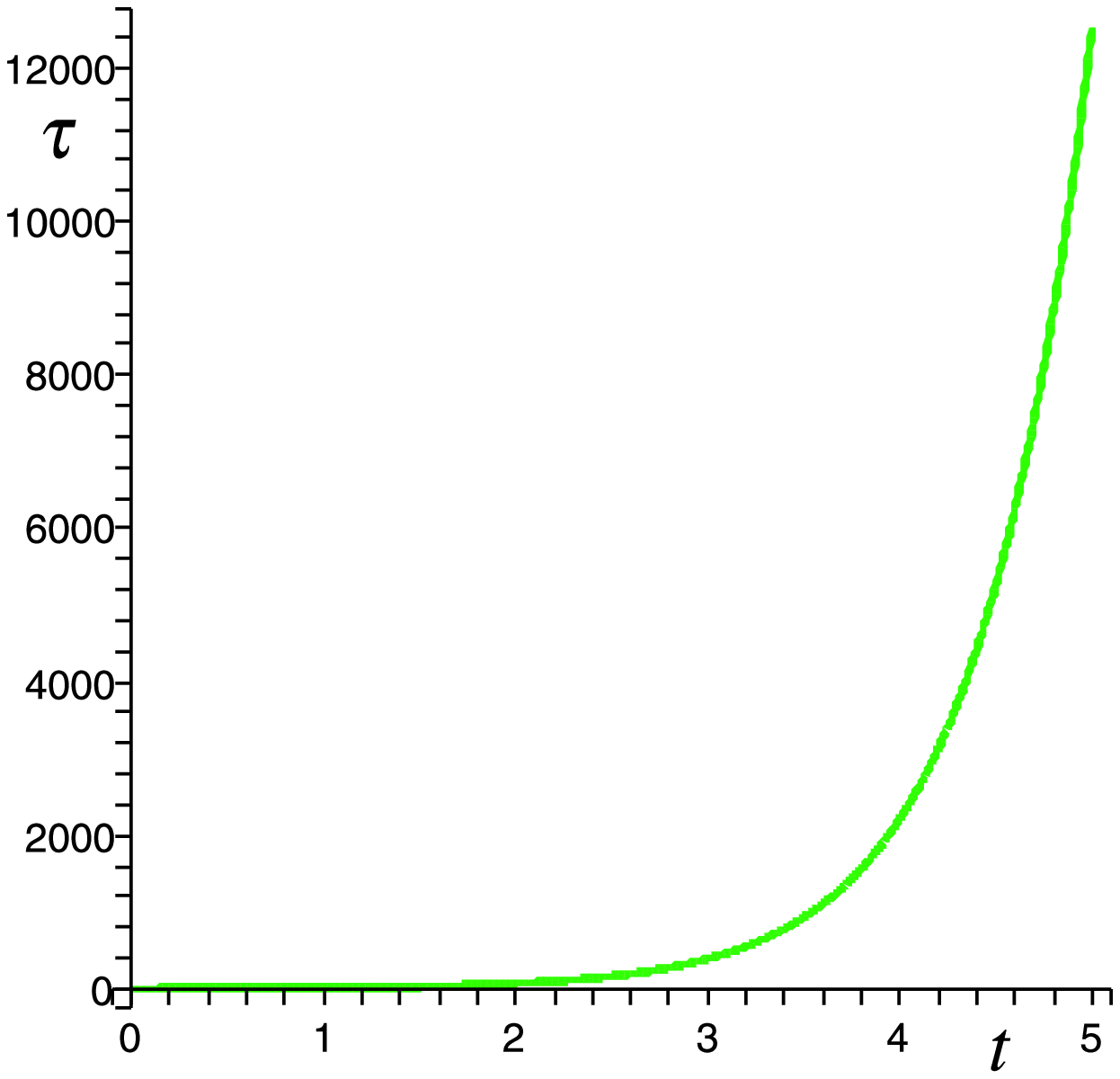}{0.45}{Evolution of $\tau$ for a positive
$\Lambda$.}{0.45} {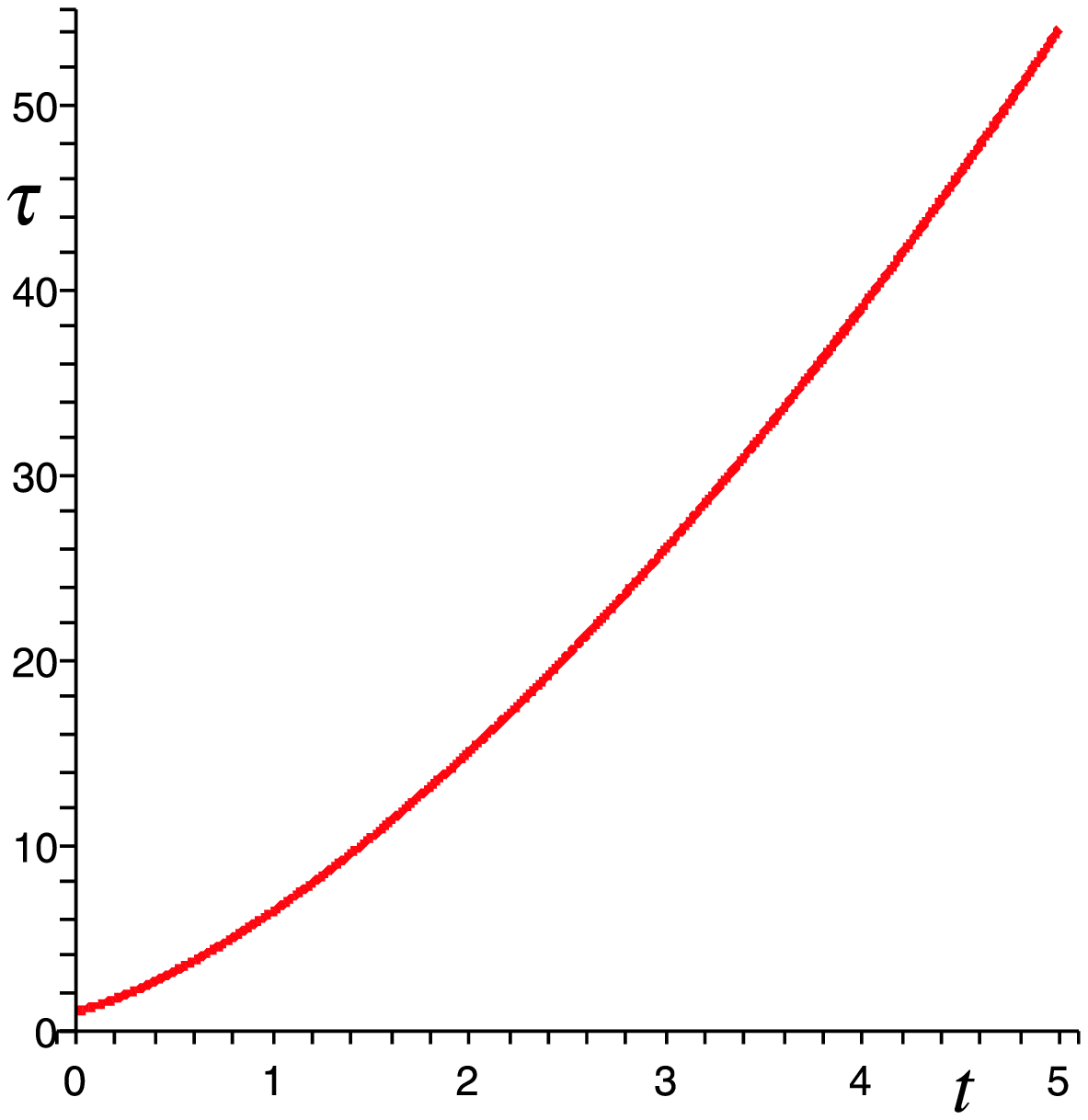}{0.45}{Evolution of $\tau$ for a trivial
$\Lambda$ term.}{0.45}

In Fig. \ref{pot1} we plot the potential corresponding to
\eqref{poten1}. In Fig. \ref{tau1m} the evolution of $\tau$ in
case of a negative $\Lambda$ is illustrated. As one sees, in this
case the model allows an oscillatory mode of expansion. In Figs.
\ref{tau1p} and \ref{tau1z} behavior of $\tau$ for a nonnegative
$\Lambda$ is presented. As one sees, introduction of a positive
$\Lambda$, which is often used to model the dark energy, results
in the rapid growth of the universe. It is evident from the Eqs.
\eqref{tau2} and \eqref{force1}, the inhomogeneity plays crucial
role at small $\tau$. In case of expanding universe
inhomogeneity becomes notable only at the early stage of
evolution, while for a oscillatory mode it is notable at $\tau =
\tau_{\rm min}$. Apparently from the expression \eqref{poten1}
varying the value of inhomogeneity parameters $m$ and $n$ the
height of the potential barrier in the vicinity of $\tau = 0$ can
be manipulated, though it will be infinitely high at $\tau = 0$.

\subsubsection{Spinor field with trigonometric induced nonlinearity}

Here we consider the case when the spinor field nonlinearity is
given by some trigonometric functions. Here we chose the following
two cases with (i) $F = \sin(S)$ and (i) $F = \exp(S)$.

If $F = \sin(S)$, we have
\begin{eqnarray}
{\cal F}(\tau,q) = 2\frac{m^2 -mn + n^2}{\tau} +
\frac{3}{2}\biggl[M  + \frac{\lambda \cos(1/\tau)}{\tau^2(1 +
\lambda \sin(1/\tau))^2}+ \frac{1-\zeta}{\tau^\zeta}\biggr] + 3
\Lambda \tau, \label{force2}
\end{eqnarray}
and
\begin{eqnarray}
{\cal U}(\tau,q) = -\{4(m^2 -mn +n^2) {\rm ln}\,\tau + 3\kappa [M
\tau -  \lambda /2(1 + \lambda \sin(1/\tau)) +  \tau^{1-\zeta}] +
3 \Lambda \tau^2\}. \label{poten2}
\end{eqnarray}

For $F = \exp(S)$ one finds
\begin{eqnarray}
{\cal F}(\tau,q) = 2\frac{m^2 -mn + n^2}{\tau} +
\frac{3}{2}\biggl[M  + \frac{\lambda \exp(1/\tau)}{\tau^2(1 +
\lambda \exp(1/tau))^2}+ \frac{1-\zeta}{\tau^\zeta}\biggr] + 3
\Lambda \tau,\label{force3}
\end{eqnarray}
and
\begin{eqnarray}
{\cal U}(\tau,q) = -\{4(m^2 -mn +n^2) {\rm ln}\,\tau + 3\kappa [M
\tau -  \lambda /2(1 + \lambda \exp(1/\tau)) +  \tau^{1-\zeta}] +
3 \Lambda \tau^2\}. \label{poten3}
\end{eqnarray}

Both these cases were solved numerically. The overall behavior of
the potential ${\cal U}$ and $\tau$ is almost the same as in case
of a power law nonlinear term illustrated in the Figs. (\ref{pot1}
- \ref{tau1z}).

\subsection{Case with $c = \sqrt{\tau}$}

In this case we find the following picture of the force ${\cal
F}(\tau,q)$ and potential ${\cal U}(\tau,q)$:
\begin{equation}
{\cal F}(\tau,q) =  2 (m^2 - mn + n^2) + \frac{3}{2} \bigl[ M +
\frac{\lambda \eta \tau^{\eta-1}}{(\lambda + \tau^\eta)^2} +
\frac{1 - \zeta}{\tau^\zeta}\bigr], \label{force4}
\end{equation}
and
\begin{equation}
{\cal U}(\tau,q) = -\{4(m^2 -mn + n^2) \tau + 3[M \tau - \lambda/
(\lambda + \tau^\eta) + \tau^{1 - \zeta}]\}. \label{poten4}
\end{equation}
Unlike the previous case now $\tau$ can be trivial as well, thus
giving rise to a spacetime singularity. Given the value of energy
level $E$ in case of a negative $\Lambda$ the solutions may be
both oscillatory and nonperiodic.

\myfigures{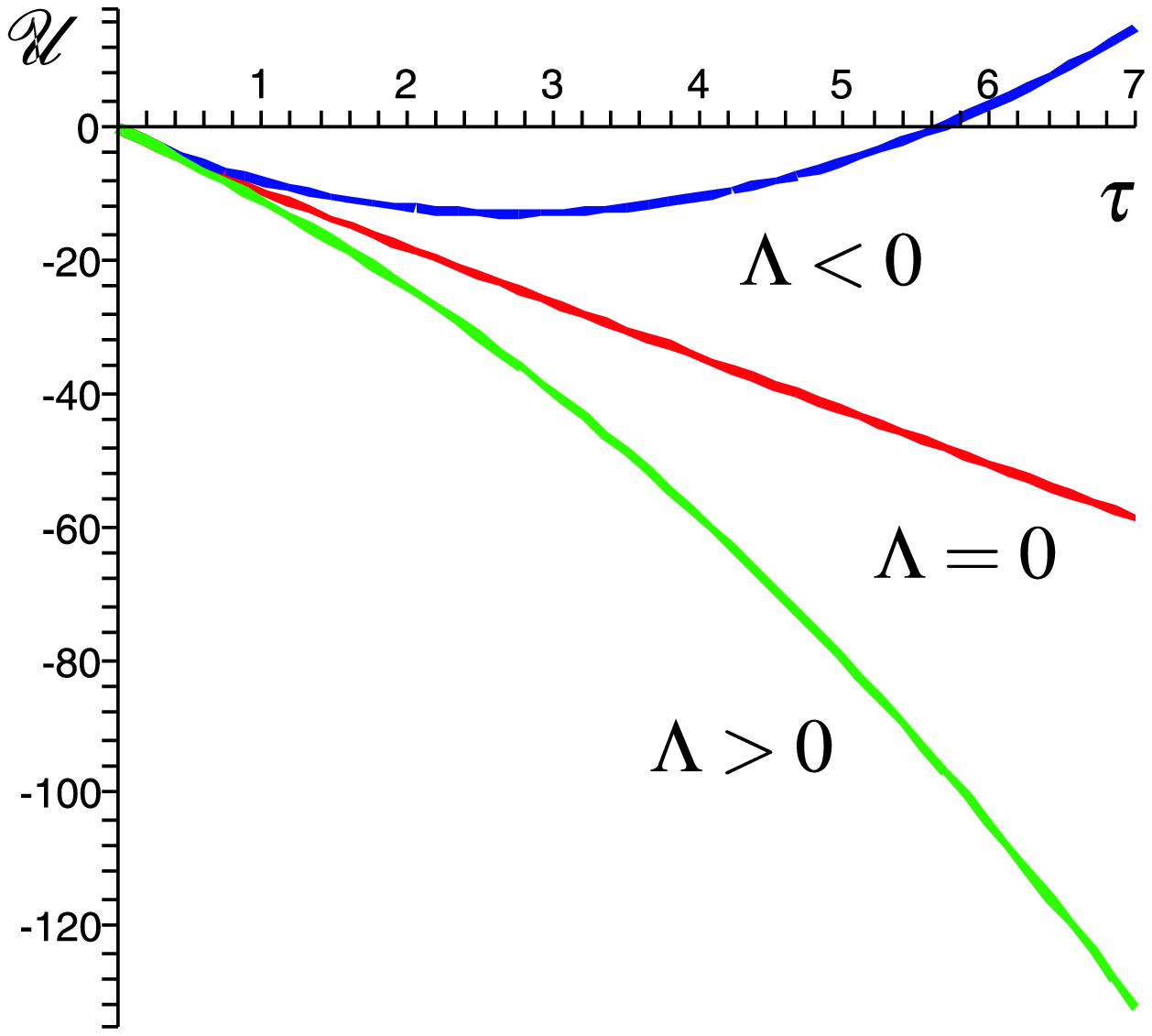}{0.45}{View of the potential ${\cal U}$ given by
\eqref{poten4} for different values of $\Lambda$.}{0.45}
{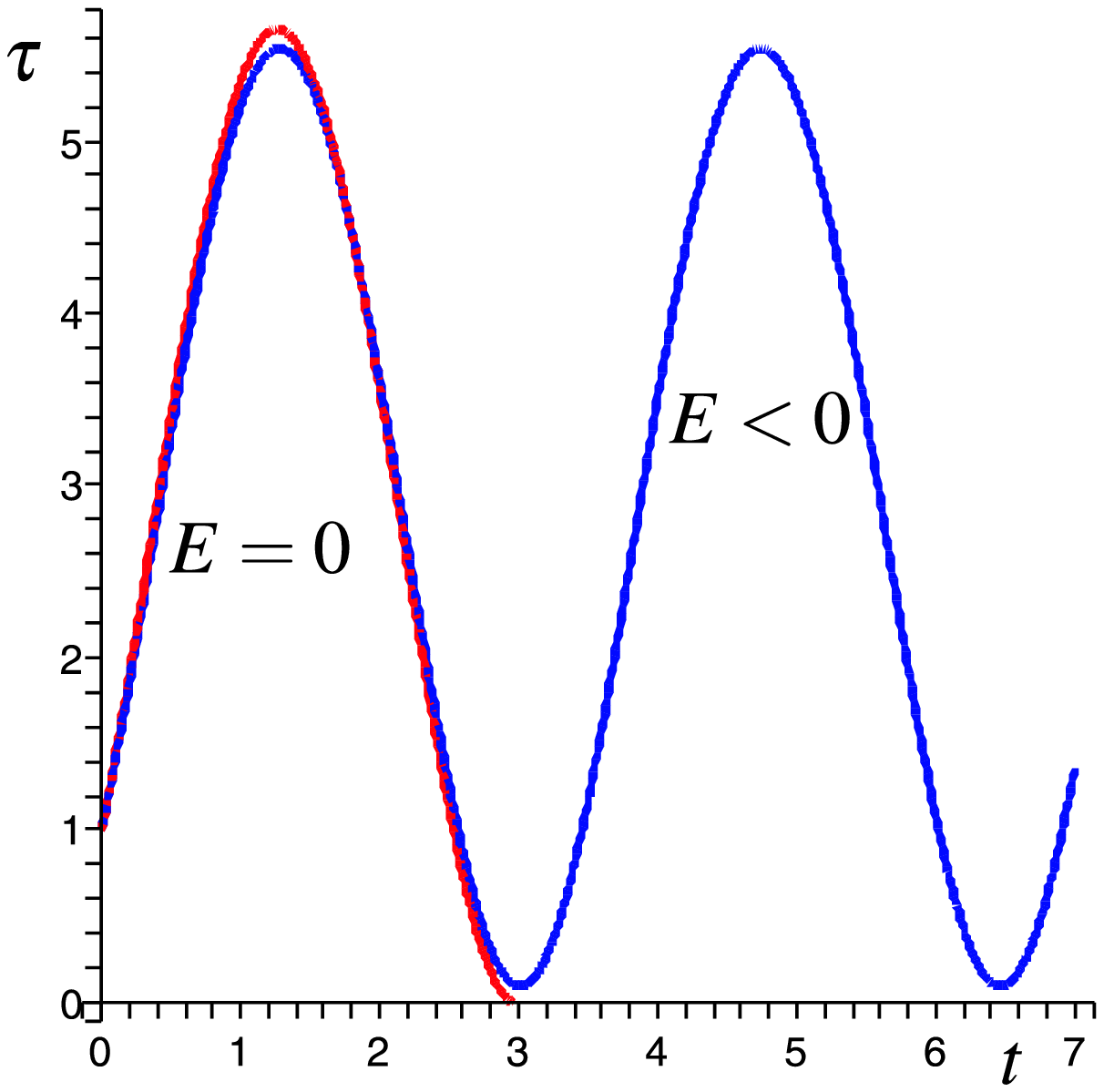}{0.45}{Evolution of $\tau$ for a negative $\Lambda$ with
different values of $E$.}{0.45}

Here as in previous case we set the following values for problem
parameters: $m=2$, $n=1$ $M=1$,$\lambda=0.1$,$\zeta=1/3$,$\eta=3$.
For cosmological constants we used $\Lambda = 0, +1, -1$,
respectively. For energy level $E$ we set $E = 0$ and $E = -1$.
The evolution of $\tau$ in case of $\Lambda \ge 0$ is same as in
previous case, while for $\Lambda < 0$ the evolution of $\tau$ is
nonperiodic for $E \ge 0$ and oscillatory for $E < 0$. In Fig.
\ref{pot4} we plot the potential corresponding to \eqref{poten4}.
Contrary to the previous case it does not possess the infinitely
high potential barrier at $\tau = 0$, which means in this case
$\tau$ can be trivial as well. In Fig. \ref{tau4n2} we illustrated
the evolution of $\tau$ for a negative $\Lambda$ with two
different values of $E$. Given the concrete value of $E$ the
solution is either oscillatory or non-periodic. As far as
inhomogeneity is concerned, in this case the part related to
inhomogeneity can be added to the mass term and it plays the same
role as the spinor mass in the evolution of the universe.

\section{Conclusion}

A self-consistent system of interacting spinor and scalar fields
within the framework of Bianchi type-VI (BVI) is studied in
presence of a cosmological constant. Exact solutions of the
spinor, scalar and gravitational field equations are obtained for
some special choice of the spinor field nonlinearity. It is shown
that introduction of a positive $\Lambda$ which is often used to
model the dark energy results in a rapid growth of the universe,
while a negative $\Lambda$ gives rise to an oscillatory or
non-periodic mode of expansion. If the metric functions $a$ and
$b$ are taken to be inverse to each other ($ab = 1$), we have a
singularity free universe independent of the sign of the $\Lambda$
term.

\clearpage

\newcommand{\hnl}{\htmladdnormallink}

\end{document}